\documentclass{aa}  

\usepackage{graphicx}
\usepackage{units}
\usepackage{lipsum}
\usepackage{txfonts}
\usepackage{graphics}
\usepackage{natbib}
\begin{document}

   \title{Quasar spectral variability from the XMM-Newton serendipitous source catalogue}

   \author{R. Serafinelli\inst{1,2} \fnmsep\thanks{roberto.serafinelli@roma2.infn.it}  \and F. Vagnetti\inst{1}\fnmsep \and R. Middei\inst{3}
          }

    \institute {Dipartimento di Fisica, Universit\`a di Roma ``Tor Vergata'', Via della Ricerca Scientifica 1, 00133, Roma, Italy
                  \and
                  Dipartimento di Fisica, Universit\`a di Roma ``La Sapienza'', Piazzale Aldo Moro 5, 00185, Roma, Italy
                  \and
                  Dipartimento di Matematica e Fisica, Universit\`a Roma Tre, Via della Vasca Navale 84, 00146, Roma, Italy}

   \date{Received XXX; Accepted YYY}

 
  \abstract
  {X-ray spectral variability analyses of active galactic nuclei (AGN) with moderate luminosities and redshifts typically show a `softer when brighter' behaviour. Such a trend has rarely been investigated for high-luminosity AGNs ($ L_{bol}\gtrsim 10^{44}$ \textrm{erg/s)}, nor for a wider redshift range (e.g. $0\lesssim z\lesssim 5$).} 
  {We present an analysis of spectral variability based on a large sample of 2,700 quasars, measured at several different epochs, extracted from the fifth release of the XMM-Newton Serendipitous Source Catalogue.}
  {We quantified the spectral variability through the parameter $\beta$ defined as the ratio between the change in the photon index $\Gamma$ and the corresponding logarithmic flux variation, $\beta=-\Delta\Gamma/\Delta\log F_X$.}
  {Our analysis confirms a softer when brighter behaviour for our sample, extending the previously found general trend to high luminosity and redshift. We estimate an ensemble value of the spectral variability parameter $\beta=-0.69\pm0.03$. We do not find dependence of $\beta$ on redshift, X-ray luminosity, black hole mass or Eddington ratio. A subsample of radio-loud sources shows a smaller spectral variability parameter. There is also some change with the X-ray flux, with smaller $\beta$ (in absolute value) for brighter sources. We also find significant correlations for a small number of individual sources, indicating more negative values for some sources.}
   {}
   {}


   \maketitle
%

\section{Introduction}
\label{sec:introduction}

All classes of active galactic nuclei (AGN) show some kind of amplitude variability, which is present in nearly all electromagnetic bands, from radio to X-rays and on several timescales. In particular, X-ray variability is observed both on very short timescales \citep{ponti12}, allowing the study of the innermost regions of the AGN, and on longer timescales of at least a number of years \citep[e.g. ][]{markowitz04, vagnetti11, shemmer14,vagnetti16}.

A geometrically thin, optically thick accretion disk around the central supermassive black hole is believed to emit optical and ultraviolet photons, while a hot corona would produce the X-rays by means of a comptonisation process of the radiation emitted by the disk itself \citep{haardt91}. While this process is not entirely understood, particularly because the shape, size and geometry of the corona are mostly unknown, variability could provide a means of procuring information about it.

Besides amplitude variability, the study of spectral variability is also interesting. This feature has been studied by several authors in the optical/UV bands \citep[e.g. ][]{giveon99, trevese02, vagnetti03} resulting in the discovery of a `harder when brighter' behaviour for the power-law spectrum, defined as $F_\nu\propto\nu^\alpha$.

In the X-ray band, spectral variability has mostly been studied for individual sources \citep[e.g. ][]{magdziarz98, mchardy01, zdziarski03,ursini16}, while few ensemble studies have been made on the topic, such as the one by \citet{sobolewska09}. The trend these studies have found is quite the opposite of the optical one, since a softer when brighter behaviour was found in both individual and ensemble studies, even at higher energies \citep{soldi14}. One notable exception is the work by \citet{connolly16}, in which the authors analysed a sample of local low-luminosity AGNs and found harder when brighter behaviour for low-Eddington-ratio sources and softer when brighter behaviour for high-Eddington-ratio sources. These observations explain the observed difference in terms of dominant seed photons source, which in the first case is cyclo-synchroton emission from the the Comptonising corona itself, while in the second case is thermal emission from the accretion disk. 

These studies, though, are mainly focused on nearby, very bright objects such as Seyfert galaxies, and very few studies were carried out concerning quasars and type-1 AGNs in a very large luminosity and redshift range. Some of these studies include the ones performed by \citet{paolillo04} and \citet{gibson12}, as well as the ones we recently published, \citet{vagnetti16} and \citet{serafinelli16}. In all these works the authors found evidence of spectral variability for some of the sources, with a softer when brighter trend. 

Section \ref{sec:dataset} briefly summarises the characteristics of our catalogue, Section \ref{sec:spvar} focuses on the detailed description of our analysis and Section \ref{sec:discussion} summarises and discusses our result.

Throughout the paper, we adopted the following cosmology: $H_0=70$ km s$^{-1}$ Mpc$^{-1}$, $\Omega_m=0.3$ and $\Omega_\Lambda=0.7$.

\section{Data}
\label{sec:dataset}

The present work makes use of the third XMM-Newton serendipitous source catalogue Data Release 5 (3XMM-DR5), in which a total of $565,962$ XMM observations of $396,910$ unique X-ray sources are listed \citep{rosen15}. This catalogue includes observations from the beginning of the mission in the year $2000$ that were made public as of the $31$st December, $2013$. All kinds of astrophysical sources are present, both galactic and extragalactic, from AGN to X-ray binaries and isolated objects such as pulsars, magnetars and X-ray emitting accreting disks around stellar-mass black holes \citep{lin12}.

In a previous paper \citep{vagnetti16}, we introduced the Multi-Epoch XMM-Newton Serendipitous AGN Sample (MEXSAS), which was obtained by cross-matching the 3XMM-DR5 catalogue with two quasar catalogues of the Sloan Digital Sky Survey, that is, SDSS-DR7Q \citep{schneider10} and SDSS-DR12Q \citep{paris16}. These two optical catalogues list $105,783$ and $297,301$ quasars,  respectively, according to slightly different definitions. The DR7Q defines a quasar as having an absolute I-band magnitude $\mathcal{M}_I\leq-22$ \citep{schneider10}, while in the DR12Q, the absolute magnitude in the I-band is $\mathcal{M}_I\leq-20.5$ \citep{paris16}, therefore the latter catalogue also lists some low-$z$ Seyfert galaxies. Moreover, there is also a subset of sources identified as Broad Absorption Line (BAL) quasars, and some radio-loud AGNs are also present (see Sec.~\ref{sec:balradio}). The cosmology used by both \citet{schneider10} and \citet{paris16} is identical to that used in this paper.

The MEXSAS catalogue has $7,837$ observations, corresponding to $2,700$ unique quasar sources. The MEXSAS is a multi-epoch catalogue, which means that a minimum of two observations for each source is required. The maximum number of observations for a single source is 39. The typical number of EPIC counts for each observation is given by $\langle \log N\rangle \sim2.2$. A summary of the number of observations and sources of the catalogues can be found in Table~\ref{tab:catalogues}.

\begin{table}
\centering
\begin{tabular}{c c c}
\hline
Catalogue & Observations & Unique sources\\
\hline
SDSS-DR7Q & $/$ & $105,783$\\
SDSS-DR12Q & $/$ & $297,301$\\
3XMM-DR5 & $565,962$ & $396,910$\\
MEXSAS & $7,837$ & $2,700$\\
\hline
\end{tabular}
\caption{Number of epochs and sources listed for each catalogue.}
\label{tab:catalogues}
\end{table}

\section{Spectral variability}
\label{sec:spvar}

To quantify the spectral variability, that is, the variations of flux with the frequency, we use the spectral variability parameter $\beta$, first introduced by \citet{trevese02}:

\begin{equation}
     \beta=\frac{\alpha(t+\Delta t)-\alpha(t)}{\log F_B(t+\Delta t)-\log F_B(t)}=\frac{\Delta\alpha}{\Delta\log F_B},
        \label{eq:betaalpha}
\end{equation}

\noindent that relates the variations of the spectral index $\alpha$, with the variations of the flux $F_B$ , integrated in a given band $B$. In the X-rays, however, it is customary to express the energy spectrum as $N(E)\propto E^{-\Gamma}$, $E=h\nu$ being the energy of the photon and $\Gamma=1-\alpha$ the photon index. Hence, we express Eq.~(\ref{eq:betaalpha}) as:
\begin{equation}
    \beta=-\frac{\Delta\Gamma}{\Delta\log F_X },
    \label{eq:betagamma}
\end{equation}

\noindent with $F_X$ being the flux in a given X-ray band.

The photon index $\Gamma$, though, is not available in our catalogue and in order to calculate Eq.~\ref{eq:betagamma} for each observation, we could either compute the spectral variability parameter by means of the hardness ratios, as described in the following section, or, alternatively, estimate the values of $\Gamma$ from the catalogue flux data.

\subsection{Ensemble analysis using hardness ratios}
\label{sec:ensembleanhr}


The hardness ratio between two given bands, $H$ and $S,$ is usually defined as:

\begin{equation}
HR=\frac{CR_H-CR_S}{CR_H+CR_S},
\label{eq:hrdef}
\end{equation}

\noindent where $CR_i$ is the count-rate in the band $i$.

The 3XMM-DR5 catalogue lists four hardness ratios: $HR_1$ is computed between the count-rates in bands $1$ and $2$, $HR_2$ between bands 2 and 3, $HR_3$ between bands 3 and 4 and finally $HR_4$ between bands 4 and 5. Band 1 includes all photons between 0.2 and 0.5~keV, while band 2 photons have energy between 0.5 and 1~keV, band 3 photons are in the range 1$-$2~keV, band 4 is the energy band 2$-$4.5~keV, and, finally, band 5 is the highest energy band, including all photons between 4.5 and 12~keV.

The hardness ratio can be related to $\beta$: 

\begin{equation}
\frac{\Delta HR}{\Delta\log F}\simeq\frac{dHR}{d\Gamma} \frac{\Delta\Gamma}{\Delta\log F}=-\beta\frac{dHR}{d\Gamma},
\label{eq:hrlogf}
\end {equation}

\noindent where we have used Eq.~\ref{eq:betagamma}. This means that the spectral variability parameter $\beta$ can be written as:

\begin{equation}
\beta=-\left( \frac{\Delta HR}{\Delta\log F} \right) \left( \frac{dHR}{d\Gamma} \right)^{-1}.
\label{eq:betahr}
\end{equation}

The denominator of Eq.~\ref{eq:betahr} can be computed analytically as a function of $\Gamma$ in the case of a spectral pure power-law model, or numerically, generating various spectral models using the \texttt{XSPEC} v.12.9.0 software package \citep{arnaud96} in cases where we want to consider a moderate absorption.

The count-rate in the band $i$ can be related to the flux by the relation $CR_i=k_i F_i$, with the coefficients $k_i$ that can be computed by the online tool \texttt{WebPIMMS}\footnote{https://heasarc.gsfc.nasa.gov/cgi-bin/Tools/w3pimmw/w3pimms.pl}. Therefore, the numerator of Eq.~\ref{eq:betahr} is approximately the linear fit between the deviations from the mean values of $HR$ and $\log F$ for a given source. Details on this procedure can be found in \citet{serafinelli16}, where various spectral models such as pure power-law and moderate absorption with different values of the column density $N_H$ have been considered, and hints of a softer when brighter behaviour were found, computing a global $\beta$ for different possible values of $\Gamma$. However, with this method, beta is derived indirectly from HR variations, and depends on Gamma through the derivative $dHR/d\Gamma$ in Eq.~\ref{eq:betahr}. It is for this reason that in this work we decided to compute the photon index $\Gamma$ directly, as described in the following section.

\subsection{Computing the photon index $\Gamma$}
\label{sec:photindex}

We now want to compute $\Gamma$ as the slope of spectral fits performed on the available spectral data, assuming a power-law behaviour for the X-ray spectrum of each observation.

The available fluxes $F_X$ are not monochromatic and are related to the monochromatic flux $F_i$ of the minimum energy of the band by the relation:

\begin{equation}
F_X = \int^{\nu_s}_{\nu_i} F_i \left( \frac{\nu}{\nu_i} \right)^{1-\Gamma} d\nu = \frac{F_i \nu_i}{2-\Gamma} \left[ \left( \frac{\nu_s}{\nu_i} \right)^{2-\Gamma} -1 \right],
\label{eq:integralflux}
\end{equation}

\noindent where $\nu_i$ and $\nu_s$ are the minimum and maximum frequencies of the band, respectively. We have also assumed a power-law spectrum of the form $F_\nu=F_i (\nu/\nu_i)^{1-\Gamma}$ over the whole X-ray band.

If we invert Eq.~\ref{eq:integralflux} and convert frequencies in energy using $E=h\nu$ , we obtain the expression for the monochromatic flux:

\begin{equation}
F_i = \frac{h F_X (2-\Gamma)}{E_i \left[ \left( \frac{E_s}{E_i} \right)^{2-\Gamma} -1 \right]},
\label{eq:monofluxes}
\end{equation}

\noindent where $h$ is the Planck constant, which in useful units of measure can be expressed as $h=4.14\cdot10^{-18} \textrm{~keV}/\textrm{Hz}$, if $E_i$ is expressed in keV.

We used Eq.~\ref{eq:monofluxes} assuming a typical photon index $\Gamma_0=1.7$, which best represents the bulk of the XMM sources, as reported by \citet{rosen15} and \citet{mateos09}. Then we computed the linear fit between $\log F_i$ and $\log E_i$ to obtain the actual photon index $\Gamma$ and its error $\sigma(\Gamma)$. In order to gain precision, we iterated using the photon indices obtained in the first step, instead of $\Gamma_0$. A further iteration did not substantially modify the results, since the convergence was reached after just two iterations.

It should be stressed that, given the strong inhomogeneity in the flux errors from band to band and even from observation to observation, and also because we are dealing with data consisting of only a small number of spectral points, the linear fits must be weighted using 

\begin{equation*}
w_i=\frac{1}{\sigma^2 (\log F_i) },
\end{equation*}

\noindent where $\sigma(\log F_i)$ is the error associated with $\log F_i$, in order to give less significance to points that deviate significantly from the power-law form but have a remarkable error.

Even though we are dealing with type-1 objects, a moderate absorption may still be present in the softer X-ray bands. However, we cannot take absorption into account since, in our analysis, we only use simple power-law spectral models. Therefore, we excluded the lowest-energy point from our spectral fits, for which the deviation from pure power-law spectrum, likely due to absorption, if present, is not negligible, and performed them with the remaining four points only. At relatively larger redshifts, this approximation works better, since the rest-frame photons are emitted at higher energies. At low redshifts, this estimation might, at most, slightly underestimate the photon indices in a few cases, however, we are mainly interested in temporal variations of $\Gamma$, thus any residual difference will be neglected.

\begin{figure}
        \includegraphics[width=\columnwidth]{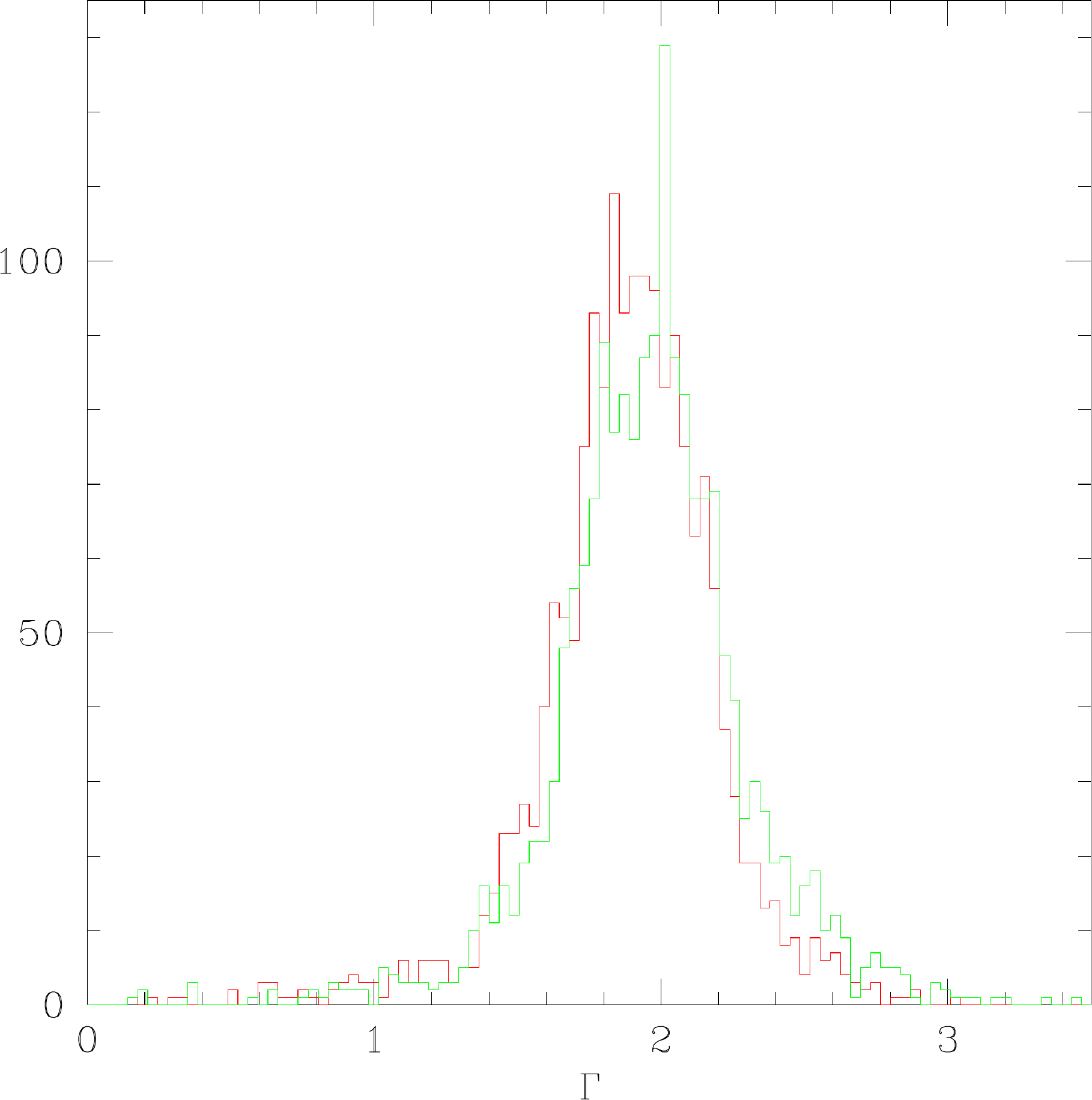}
    \caption{The green histogram is the distribution of \texttt{WAPO\_GAMMA} in the coincident data between our catalogue and the XMMFITCAT catalogue. The red histogram is the distribution of $\Gamma$ computed for the same subsample with our method.}
    \label{fig:gammadistr}
\end{figure}

In order to check the validity of this method, we confronted our $\Gamma$s with the XMMFITCAT spectral catalogue\footnote{http://xraygroup.astro.noa.gr/Webpage-prodec/thecatalog.html, recently updated to include data from 3XMM-DR5.} by \citet{corral15}, in which the authors computed the $\Gamma$s of a limited subset with a high number of counts ($N\geq50$) of the 3XMM-DR5 catalogue. We created a subsample of coincident data with the XMMFITCAT catalogue, which consists of 2,002 observations of 707 sources, and we compared our $\Gamma$s in this subsample with the \verb|WAPO_GAMMA| entry, corresponding to a power-law with intrinsic absorption spectral fit of the XMMFITCAT catalogue (for this matched subsample, the number of counts is $N\gtrsim100$). Not all the observations are well fitted by an absorbed power-law spectrum, and \citet{corral15} flag those observations that are well fitted by a particular spectral model, requiring the \texttt{XSPEC} command \texttt{goodness} to return a value lower than 50\%. Since 90\% of the observations satisfy this requirement, we only compare our $\Gamma$ with the ones from this subset, which consists of 1,771 observations of 630 unique sources. As can be seen in Fig.~\ref{fig:gammadistr}, the two $\Gamma$ distributions are very similar (with a small shift between the mean values $\langle\texttt{WAPO\_GAMMA}\rangle-\langle\Gamma\rangle \simeq0.07$), suggesting that our method is robust, even if the $\Gamma$s are obtained from only four points instead of from a whole spectrum.

We stress that our pure power-law approximation could not properly represent the actual spectral shape, in some cases affected by the presence of additional components, such as soft excess \citep[e.g. ][]{turner88} or complex absorption, such as the warm absorber \citep[e.g. ][]{reynolds95}, for example. The incidence of these cases is likely limited to approximately 10\% of the observations, as in the XMMFITCAT subsample. Moreover, we assume that the effect of these phenomena is statistically negligible for the estimate of the $\Gamma$ variations, which are mainly involved in our subsequent analysis (see Sect.~\ref{sec:photindan}).

\begin{figure}
        \includegraphics[width=\columnwidth]{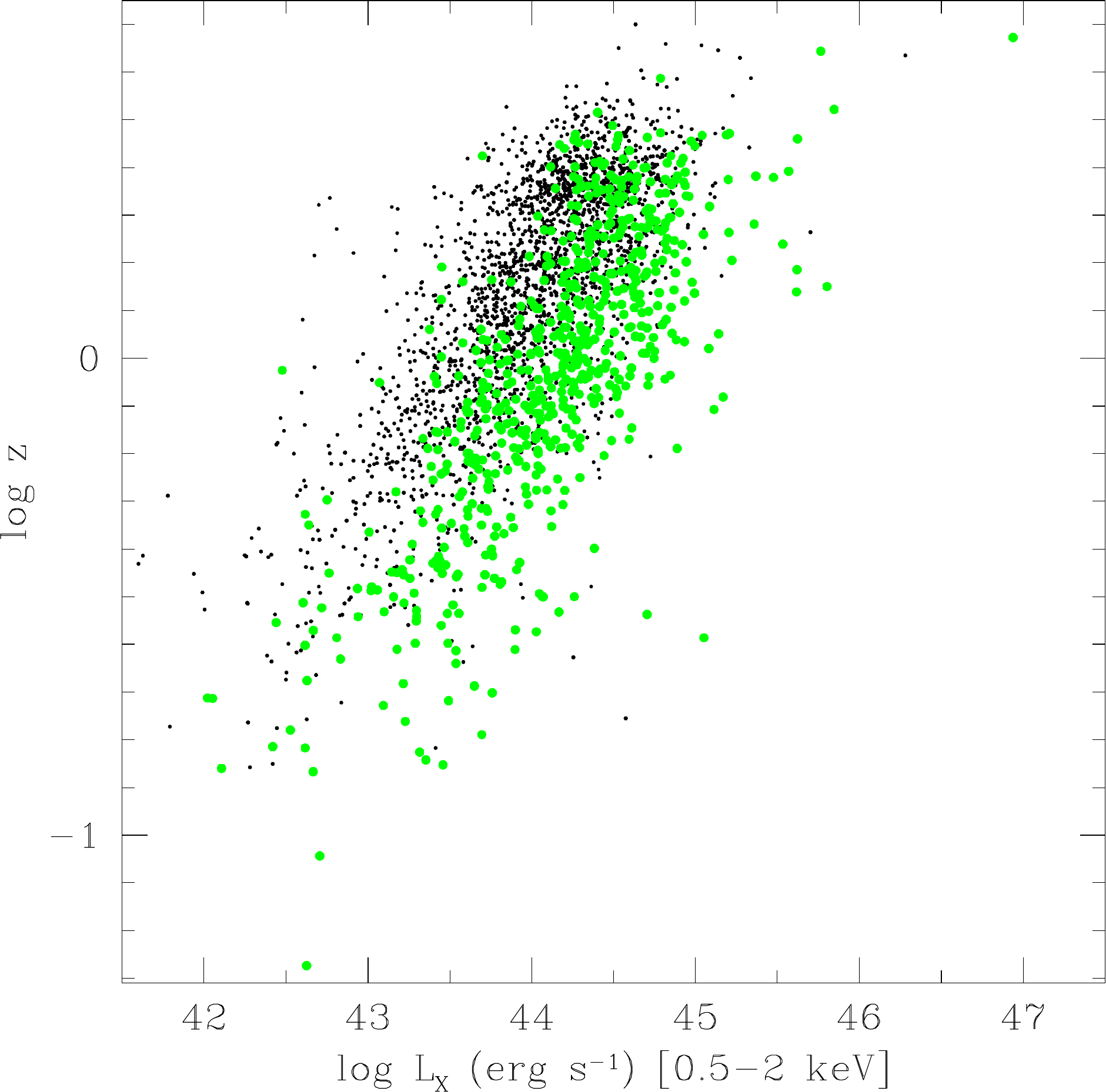}

        \includegraphics[width=\columnwidth]{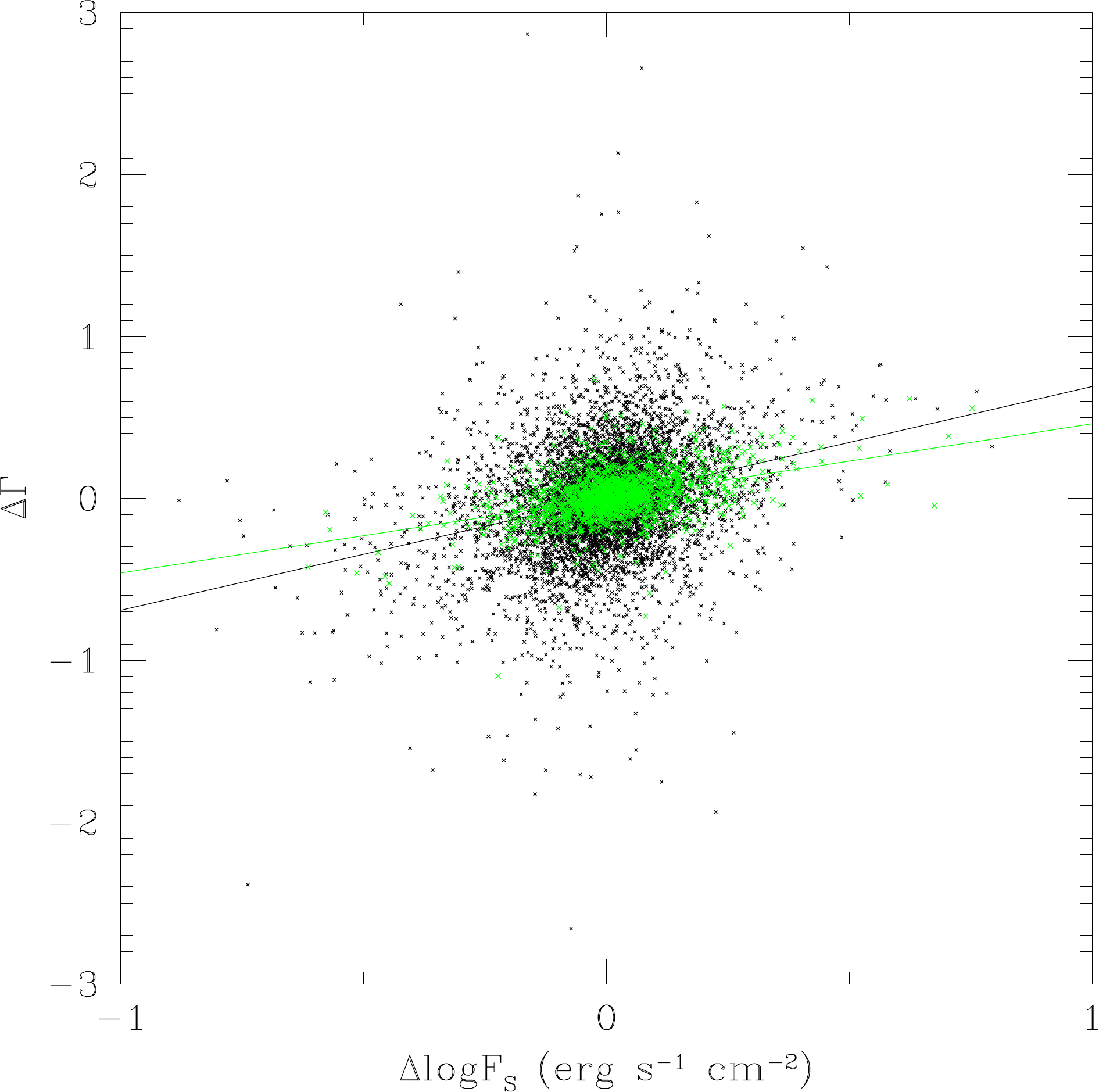}
    \caption{\textit{Top}: relation between the X-ray luminosity $L_X$ and the redshift $z$. Sources in green are the ones that are present in both the XMMFITCAT and MEXSAS catalogue, while the black dots are sources only present in the latter. The green circles are selected by \citet{corral15}, when the number of counts is at least 50, and these sources are, on average, brighter than the ones not included. \textit{Bottom}: $\Delta\Gamma-\Delta\log F_S$ correlation, plotted using the photon indices from our analysis, for the same two subsets. The green line shows the linear fit of just the green data, while the black line is the linear fit of the whole sample.}
    \label{fig:lxz}
\end{figure}

\begin{figure}
        \includegraphics[width=\columnwidth]{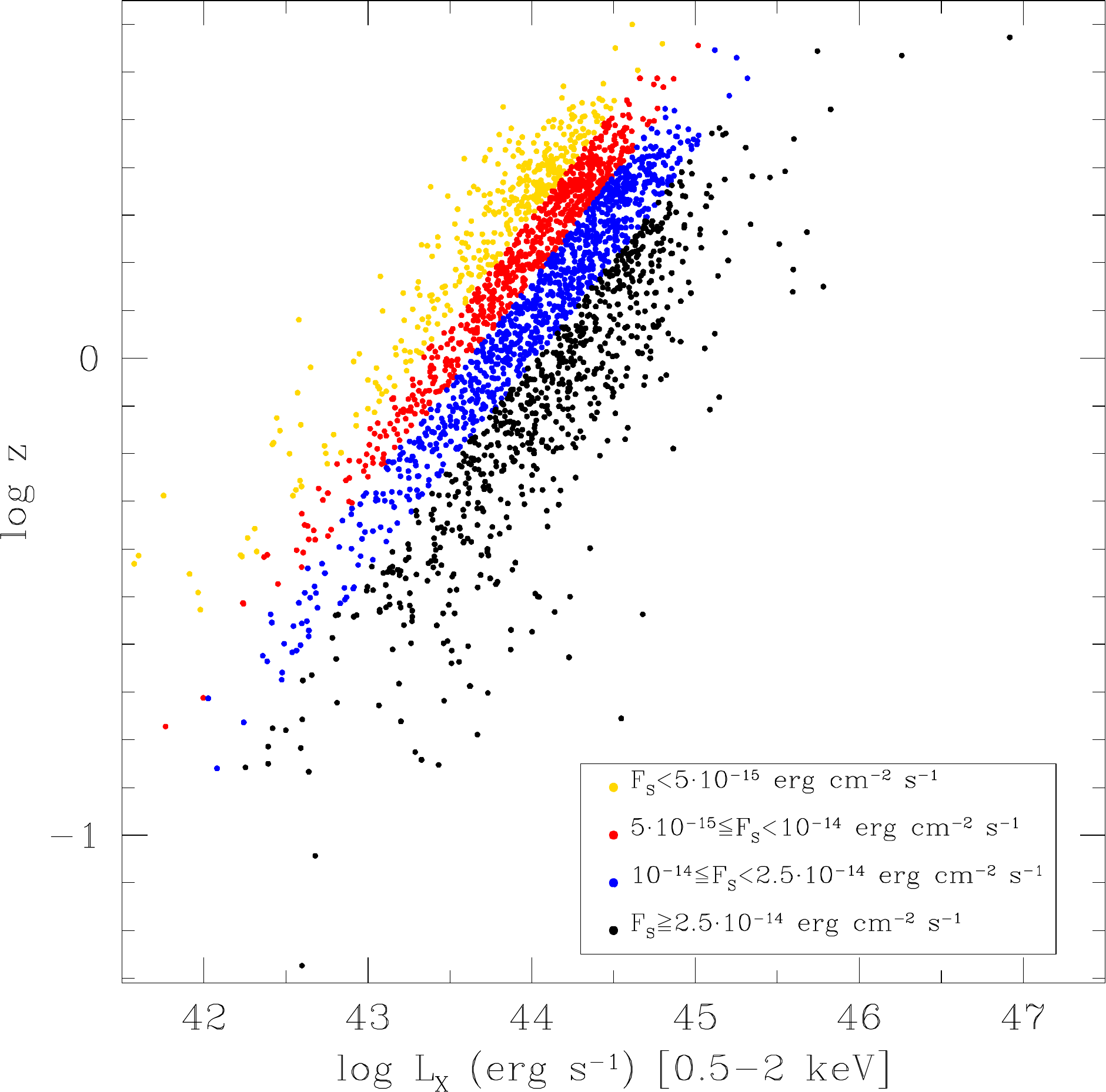}

        \includegraphics[width=\columnwidth]{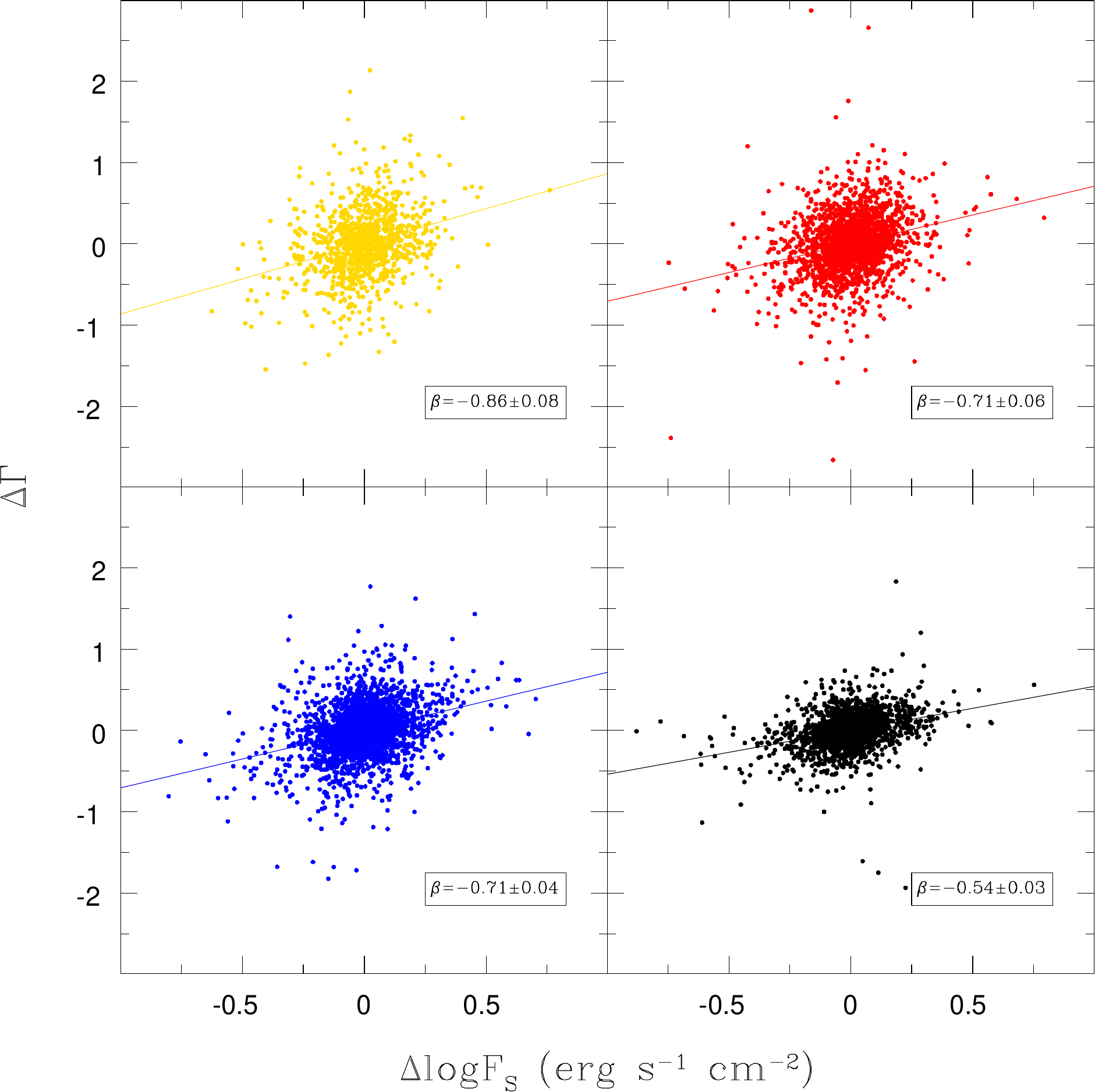}
    \caption{\textit{Top}: distribution of the sources in the $z-L_X$ plane. Positions of four different flux bins in the $z-L_X$ plane are clearly distinguishable as parallel stripes. Different colours represent different average flux bins. \textit{Bottom}: $\Delta\Gamma-\Delta\log F_S$ plots for the same four flux bins, drawing the linear fit lines for each bin. The slope, which is the opposite of $\beta$, decreases for increasing fluxes. The correlation coefficients are approximately $0.3$ and are reported in Table~\ref{tab:betaflux}. All probabilities of obtaining such correlations by chance are $p(>r)<10^{-10}$.}
    \label{fig:dgdlogf1}
\end{figure}

\subsection{Ensemble analysis using the computed photon indices}
\label{sec:photindan}

Once the photon indices have been calculated for all the observations in our catalogue, in principle one should be able to compute the spectral variability parameter $\beta$ using Eq.~\ref{eq:betagamma}; that is, we are trying to find a correlation between $\Delta\Gamma$ and $\Delta\log F$.

The flux chosen for the search of such correlation is not any of those listed in the catalogue. Many authors (e.g. \citet{paolillo04}) use typical soft ($0.5-2$ keV) and hard ($2-10$ keV) bands, therefore we have computed the integrated flux in the band $0.5-2$ keV, which is the sum of the bands 2 and 3; we refer to this as $F_S$. Following the instructions of \citet{watson09}, the soft flux $F_S$ is given by:

\begin{equation}
F_S = W_{pn,S}F_{pn,S}+W_{m1,S}F_{m1,S}+W_{m2,S}F_{m2,S},
\label{eq:sumfluxes}
\end{equation}

\noindent where $F_{pn,S}$, $F_{m1,S}$ and $F_{m2,S}$ indicate the fluxes received in the soft band S by the three X-ray cameras of the detector. The weights are in turn given by: 

\begin{equation}
W_{pn,S}=\frac{ \sigma^{-2 }\left (F_{pn,S}\right )  }{\sigma^{-2} \left (F_{pn,S}\right )+\sigma^{-2} \left (F_{m1,S}\right )+\sigma^{-2} \left (F_{m2,S} \right )},
\label{eq:weightfluxes}
\end{equation}

\noindent where $\sigma(F_i)$ indicates the error associated with a given flux $F_i$. 

For a given individual source, the opposite of the slope obtained by a linear fit between $\Gamma$ and $\log F$ is a good approximation of the spectral variability parameter $\beta$. This is not true for an ensemble analysis, in which the average photon indices and fluxes might be very different from source to source. To increase coherence between different sets of observations from different sources, instead of simply computing the linear fit between $\Gamma$ and $\log F_S$, we have computed such linear fits between the deviations of $\Gamma$ from the source mean value $\overline{\Gamma}$, as well as the deviations of the logarithm of the flux from its mean value $\overline{\log F}_S$. The slope of this linear fit gives 

\begin{equation}
\beta=-0.69\pm0.03.
\label{eq:betaensemble}
\end{equation}

\begin{figure}
        \includegraphics[width=\columnwidth]{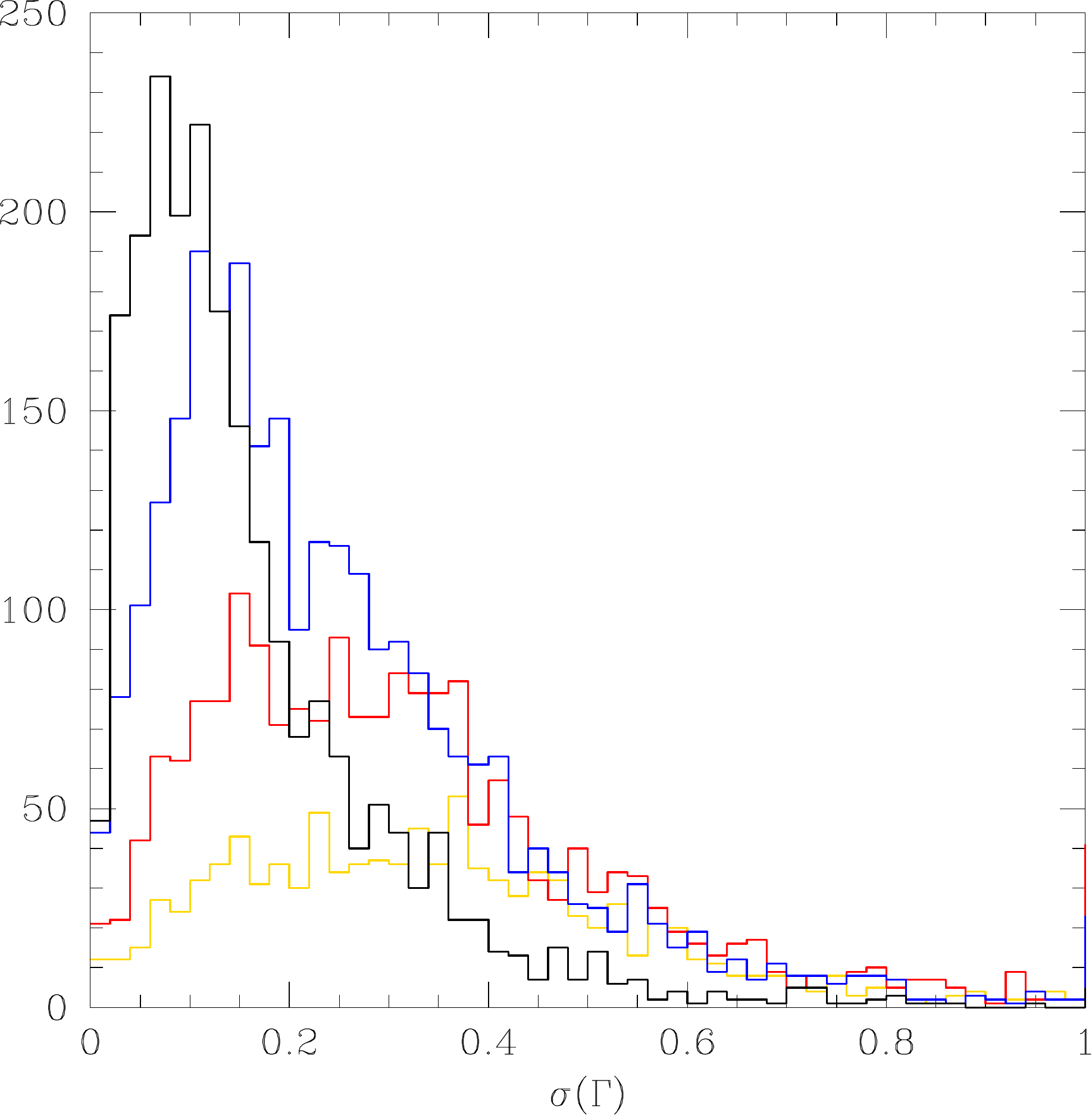}
       
        \includegraphics[width=\columnwidth]{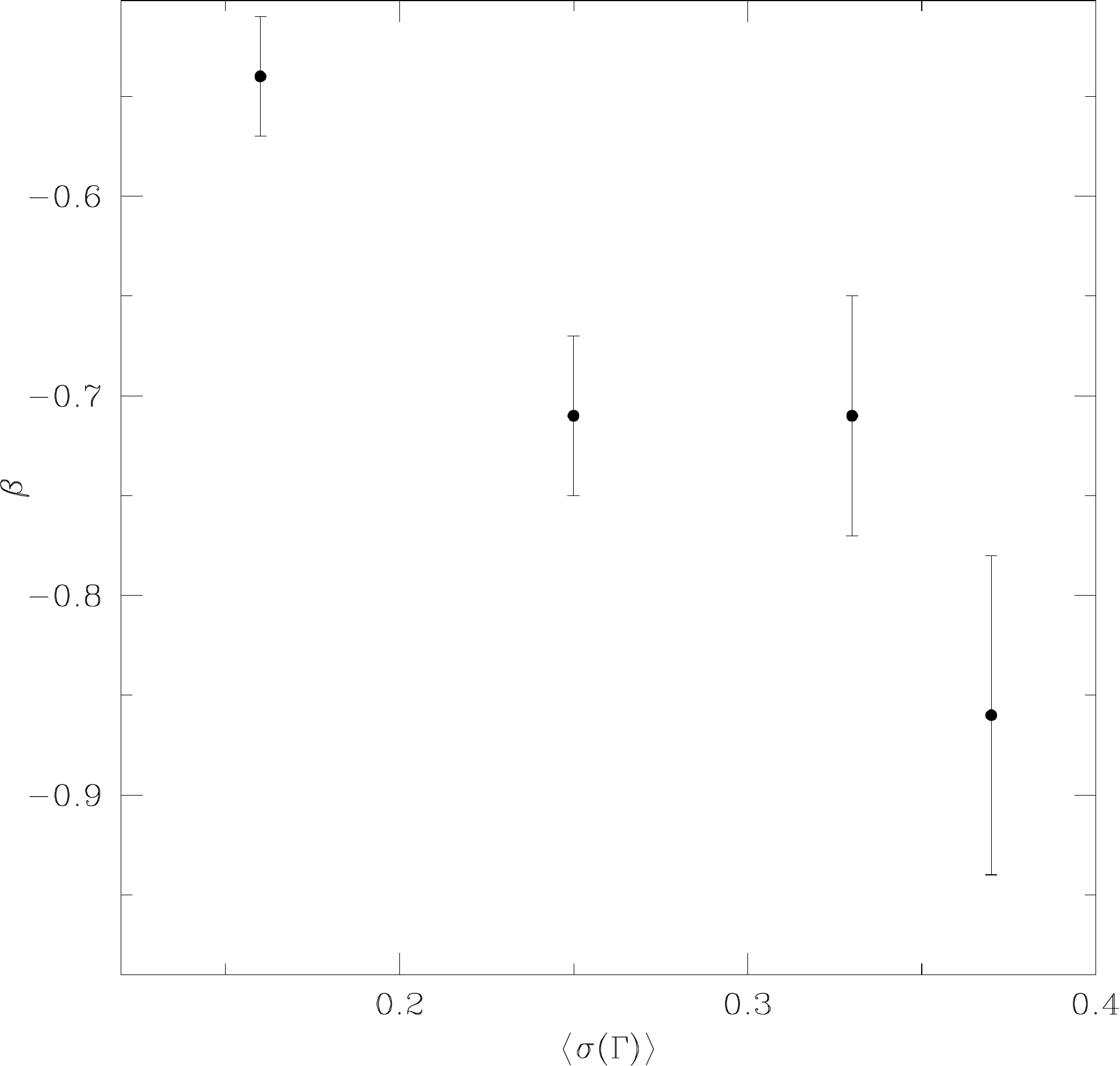}
    \caption{\textit{Top}: $\sigma(\Gamma)$ distribution for the same flux bins defined in Fig.~\ref{fig:dgdlogf1}. Bins of higher fluxes have smaller average errors and viceversa.  \textit{Bottom}: at smaller average $\Gamma$ errors, we have smaller absolute values of $\beta$ for the corresponding bins.}
    \label{fig:dgdlogf2}
\end{figure}

The correlation coefficient is $r=0.293$, with negligible probability of finding this correlation by chance, $p<10^{-10}$. Given the high heterogeneity of the sample, such moderate correlation and slope in the whole sample are not surprising. The small value of the probability indicates that this correlation, although small, is significant.

In order to give better significance to the data, we also defined a better subsample, in which the worst observations are eliminated, meaning we removed the observations with $\sigma\left (\Gamma\right )>0.5$. Since we are performing spectral fits using only four points, we are likely to obtain significant errors on $\Gamma$. This subsample discards 1,245 observations, leaving us with 6,592 observations of 2,309 unique sources. Since we found a value of $\beta=-0.68\pm0.02$ for this subsample, no significant difference on the ensemble $\beta$ was found, and therefore we kept working on the whole sample.

To further validate the consistency of our computing method for the photon indices $\Gamma$ (see Sec.\ref{sec:photindex}), we computed $\beta$ for the coincident subsample between the XMMFITCAT and our catalogue using both our $\Gamma$ and the \texttt{WAPO\_GAMMA} entry, representing the $\Gamma$s obtained by spectral fits assuming an absorbed power-law behaviour of the spectra. We find $\beta=-0.46\pm0.02$ for our $\Gamma$ values and $\beta=-0.46\pm0.04$ for the $\Gamma$ values that are present in the XMMFITCAT catalogue. This is further evidence that the calculation of $\Gamma$ with our method is a good approximation of the actual photon index.

This value is smaller than the one obtained by the whole sample. As is clear from Fig.~\ref{fig:lxz}, the XMMFITCAT sample is made up of brighter sources, since they have better spectra. We divided the sample into four bins of the average flux of each source, which are clearly distinguishable in the $z-L_X$ plane, and this means that, on average, the spectrum of brighter sources appears to be less variable than the spectrum of fainter ones. To show this, we computed $\beta$ for each bin, obtaining a spectral variability parameter that approximately decreases in absolute value with higher average fluxes (see Fig.~\ref{fig:dgdlogf1} and Table \ref{tab:betaflux}). This is due to the average higher error in the $\Gamma$s of fainter flux bins (see Fig.~\ref{fig:dgdlogf2}).

We stress that these ensemble $\beta$s are average values that show how the spectra of these sources vary, on average, with flux. As we see in Sec.\ref{sec:indisources}, individual sources may even significantly differ from these average values. The negative ensemble value of $\beta$ indeed confirms a global softer when brighter behaviour, confirming the preliminary result obtained with the hardness ratio method. However, this result does not exclude non-negative values of $\beta$, meaning a harder when brighter behaviour for some individual sources, which are indeed present in the sample. We did not, however, find any such sources among the most significantly correlated ones (see Sect.\ref{sec:indisources}). In our previous paper \citep{vagnetti16}, we estimated the ensemble $\beta$ of the whole MEXSAS catalogue by means of the structure function, and found the value $\beta=-0.35\pm0.02$. Even though the computation of $\beta$ with the structure function is an indirect method, which was performed using just part of the SF ($100-1000$ days) and only the three softest XMM-Newton energy bands, that estimate agrees with the general trend found in the present analysis.

\begin{table*}
\begin{center}
\begin{tabular}{c c c c c c c }
\hline
$\langle F_S \rangle_{min}$ & $\langle F_S \rangle_{max}$ & $N_{sour}$ & $N_{obs}$ & $\beta\pm\sigma(\beta)$ & $\langle\sigma(\Gamma)\rangle$ & $r$\\
\hline
$4.26\cdot10^{-16}$ & $5\cdot10^{-15}$ & $401$ & $1039$ & $-0.86\pm0.08$ & $0.37$ & $0.295$\\
$5\cdot10^{-15}$ & $10^{-14}$ & $729$ & $1923$ & $-0.71\pm0.06$ & $0.33$ & $0.261$\\
$10^{-14}$ & $2.5\cdot10^{-14}$ & $963$ & $2689$ & $-0.71\pm0.04$ & $0.25$ & $0.317$\\
$2.5\cdot10^{-14}$ & $4.30\cdot10^{-12}$ & $715$ & $2186$ & $-0.54\pm0.03$ & $0.16$ & $0.329$\\
\hline
\end{tabular}
\caption{List of four $\langle F_S \rangle$ flux bins with $\beta\pm\sigma(\beta)$, number of sources $N_{sour}$, number of observations $N_{obs}$ and correlation coefficient $r$. A column showing the average $\Gamma$ error for each bin is also shown, from which it can be seen that fainter sources also have average greater errors on the spectral slope $\Gamma$. The bins at higher fluxes have a smaller $\beta$ (in absolute value), which means they are less variable in spectrum, but higher correlations are found for these sources. All probabilities of finding the correlations by chance are found to be smaller than $10^{-10}$.}
\label{tab:betaflux}
\end{center}
\end{table*}

\subsection{Dependence on physical parameters}
\label{sec:subsamples}

As described in Sect.~\ref{sec:photindan}, our sample is highly heterogeneous. In order to obtain a more detailed analysis, we divided our entire sample into bins of given variables such as redshift, X-ray luminosity, black hole mass and Eddington ratio.

This section demonstrates our method of finding for which values of any one of these given variables the spectrum deviates more or less from the ensemble value, in order to find any possible dependence of the spectral variability parameter on any of these quantities.

\subsubsection{Redshift}
\label{sec:binredshift}

\begin{figure}
\includegraphics[width=\columnwidth]{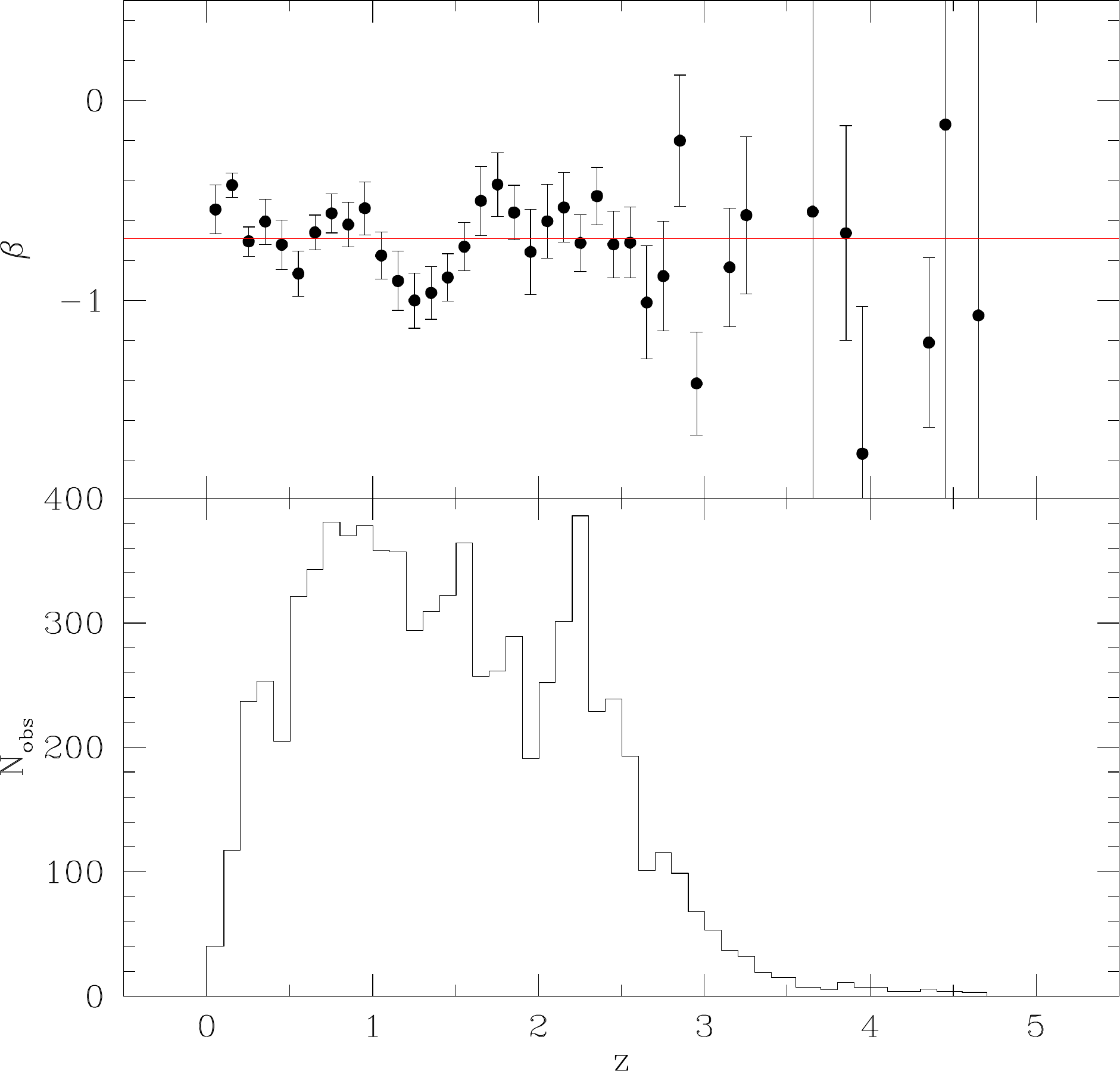}
\caption{\textit{Top}: dependence of $\beta$ on the redshift. The red line represents the ensemble value $\beta=-0.69$. The value of $\beta$ doesn't show a clear trend with redshift, although some deviations are present for some bins. \textit{Bottom}: redshift histogram of the sample. Up to $z\simeq2.8$, the number of points in each bin is $\gtrsim10^2$, while for $z\gtrsim2.8$ the bins include much fewer points, and therefore the $\beta$s have greater errors.}
\label{fig:betaredshift}
\end{figure}

Both SDSS quasar catalogues DR7Q and DR12Q include the redshift of each source, which is then available for all the sources in our sample.

 We divided the whole sample into redshift bins, each containing all the points within a redshift range of size $\delta z=0.1$ . For each bin, we have computed $\beta$ using Eq.~\ref{eq:betagamma}.

As can be seen in the upper plot of Fig.~\ref{fig:betaredshift}, there is no clear dependence of $\beta$ on the redshift. As shown in the bottom plot of Fig.~\ref{fig:betaredshift}, for $z\gtrsim2.8$ the number of points drops below $\simeq 10^2$. Since the fits performed to compute $\beta$ are less reliable with decreasing number of points, the errors on $\beta$ are higher.

Thus, we can safely assume that $\beta$ does not depend on the redshift, therefore the mechanism behind it is not related to the redshift of the source.

\begin{figure}
\includegraphics[width=\columnwidth]{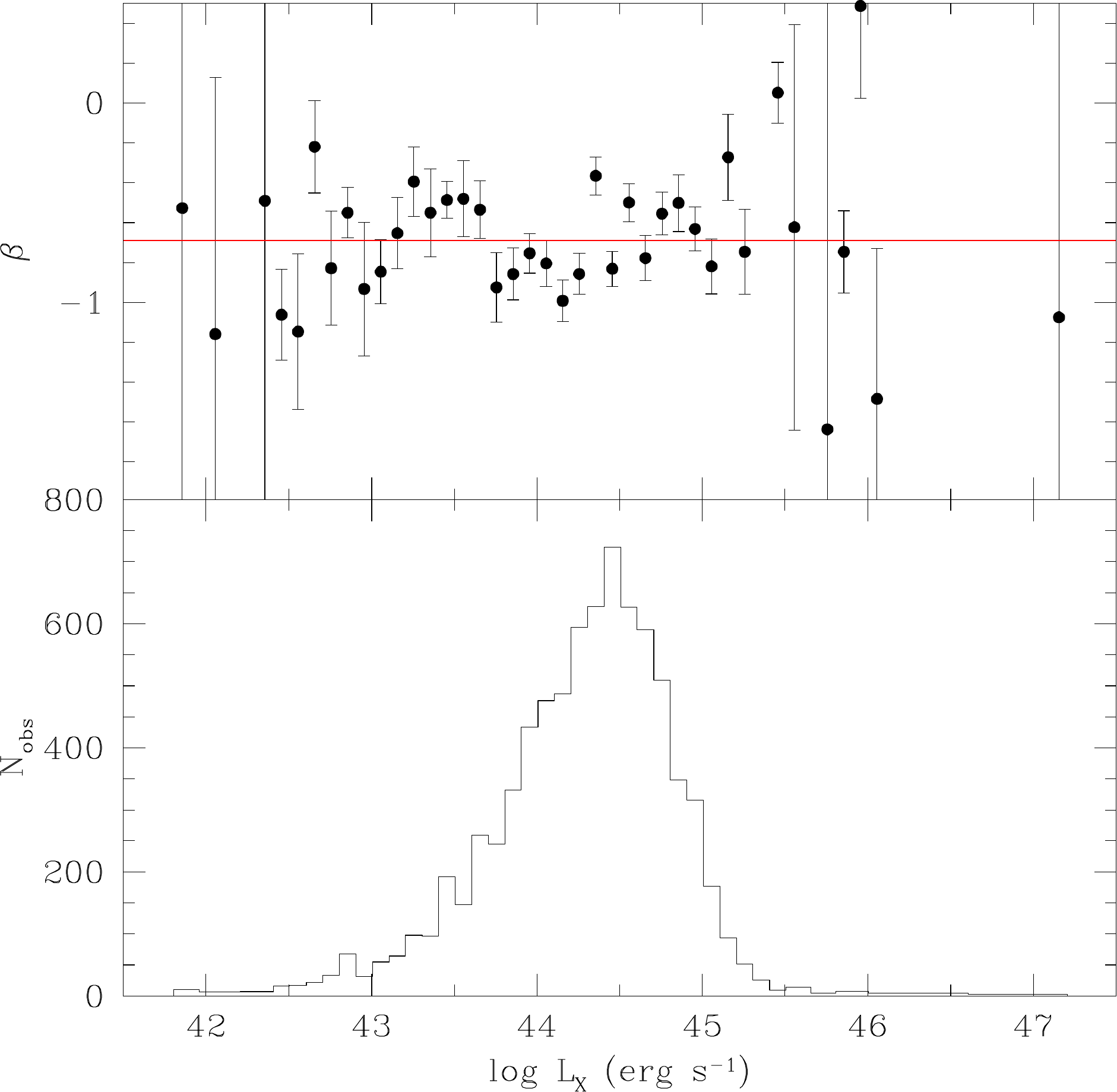}
\caption{\textit{Top}: spectral variability parameter $\beta$ as a function of the X-ray luminosity. The red line represents the ensemble value $\beta=-0.69$. \textit{Bottom}: histogram of X-ray luminosity. The errors on $\beta$ are very large in those $L_X$ intervals where the histogram has $\lesssim 10^{2}$ points.}
\label{fig:betalx}
\end{figure}

\subsubsection{X-ray luminosity}
\label{sec:binxlum}

It is well known that there is a strong observational correlation between the redshift $z$ and the X-ray luminosity $L_X$. The location of the sources in the luminosity-redshift plane is shown in Fig.~\ref{fig:lxz}.

None of the catalogues we used include a list of X-ray luminosities for each epoch, therefore we had to compute it. Using the cosmology introduced in Sect.\ref{sec:introduction}, we compute the luminosity distance of each source as: 

\begin{equation*}
D_L(z)=\frac{c}{H_0}(1+z) \int^z_0{\frac{dz}{\sqrt{\Omega_m \left( 1+z \right)^3+\Omega_\Lambda}}},
\end{equation*}

\noindent and then we compute the X-ray luminosity as:

\begin{equation}
L_X=4\pi F_X D_L^2(z)\left( 1+z \right)^{\Gamma-2},
\label{eq:lxz}
\end{equation}

\noindent where we have used the $F_S$ integrated flux between 0.5 keV and 2 keV for the X-ray flux $F_X$, the redshift of each source and we have adopted a mean value $\Gamma=1.7$ for every point. The mean value of $L_X$ was then computed for each source.
 The choice of using a fixed average $\Gamma$ was taken mostly because the $\Gamma$s we are computing are affected by relatively large errors. While the variations $\Delta\Gamma$, and therefore the computation of $\beta$, are less affected by these errors, the use of a mean value is preferred for the computing of $L_X$.

Once we have defined the X-ray luminosity of each source, we grouped the observations in X-ray luminosity bins of size $\delta \log L_X = 0.1$ and computed $\beta$ for each bin, using the same procedure adopted for the redshift. The results are shown in Fig.~\ref{fig:betalx}, showing, again, no dependance of $\beta$ on the X-ray luminosity.

This should not be surprising, since $z$ and $L_X$ are not independent quantities, as they are strongly related (see Fig.~\ref{fig:lxz}). However, it does prove that the spectral variability, and hence the mechanism behind it, does not depend on the X-ray luminosity of the source.

\begin{figure}
\centering
\includegraphics[width=\columnwidth]{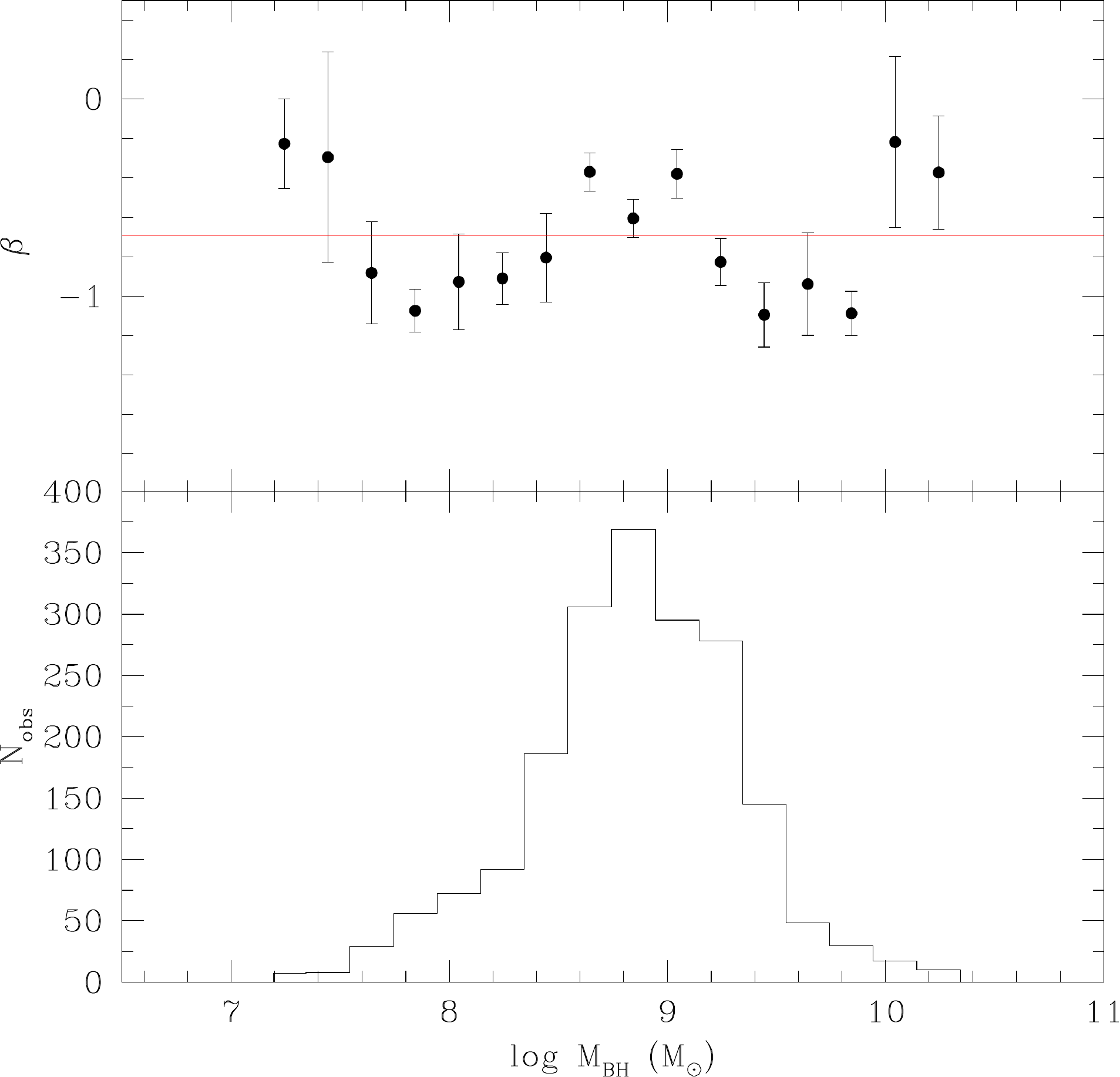}
\caption{\textit{Top}: spectral variability parameter $\beta$ versus black-hole mass $M_{BH}$. Aside from statistical fluctuations, no clear trend is found between the black-hole mass and $\beta$. \textit{Bottom}: histogram for black-hole mass.}
\label{fig:bhmass}
\end{figure}

\subsubsection{Black-hole mass and Eddington ratio}
\label{sec:bhedd}
The MEXSAS catalogue, being derived from XMMSSC-DR5 and from the Sloan quasar catalogues, does not include distinguishing parameters of the source other than the redshift. Moreover, as described in Sect.~\ref{sec:binxlum}, we computed the average X-ray luminosity of each source.

To study further parameters, such as the central-black-hole mass and the Eddington ratio of the source, we matched the MEXSAS catalogue with a catalogue containing the properties of 105,783 quasars \citep{shen11}. This catalogue\footnote{http://quasar.astro.illinois.edu/BH\_mass/dr7.htm} is limited, however, because it only includes sources that are present on SDSS-DR7Q, and therefore crossing it with MEXSAS means we did not analyse all those sources that are present in the DR12Q but not in the DR7Q. We obtain a subsample made up of 565 unique sources and 1,953 observations.

First, we divided this sample into bins of black-hole mass logarithm of $\delta\log M_{BH}=0.2$ amplitude and computed the ensemble $\beta$ for each bin. As we can see in Fig.~\ref{fig:bhmass}, aside from statistical fluctuations from the ensemble value $\beta=-0.69$, a clear trend is not found, and this means that neither the black-hole mass is a driver parameter for the spectral variability.

It should be stressed that the black-hole masses listed in \cite{shen11} are single epoch virial estimates computed from the H$\beta$, MgII and CIV lines, depending on the redshift of the source. These estimates, especially the masses obtained from the CIV line, are highly uncertain.

\begin{figure}
\includegraphics[width=\columnwidth]{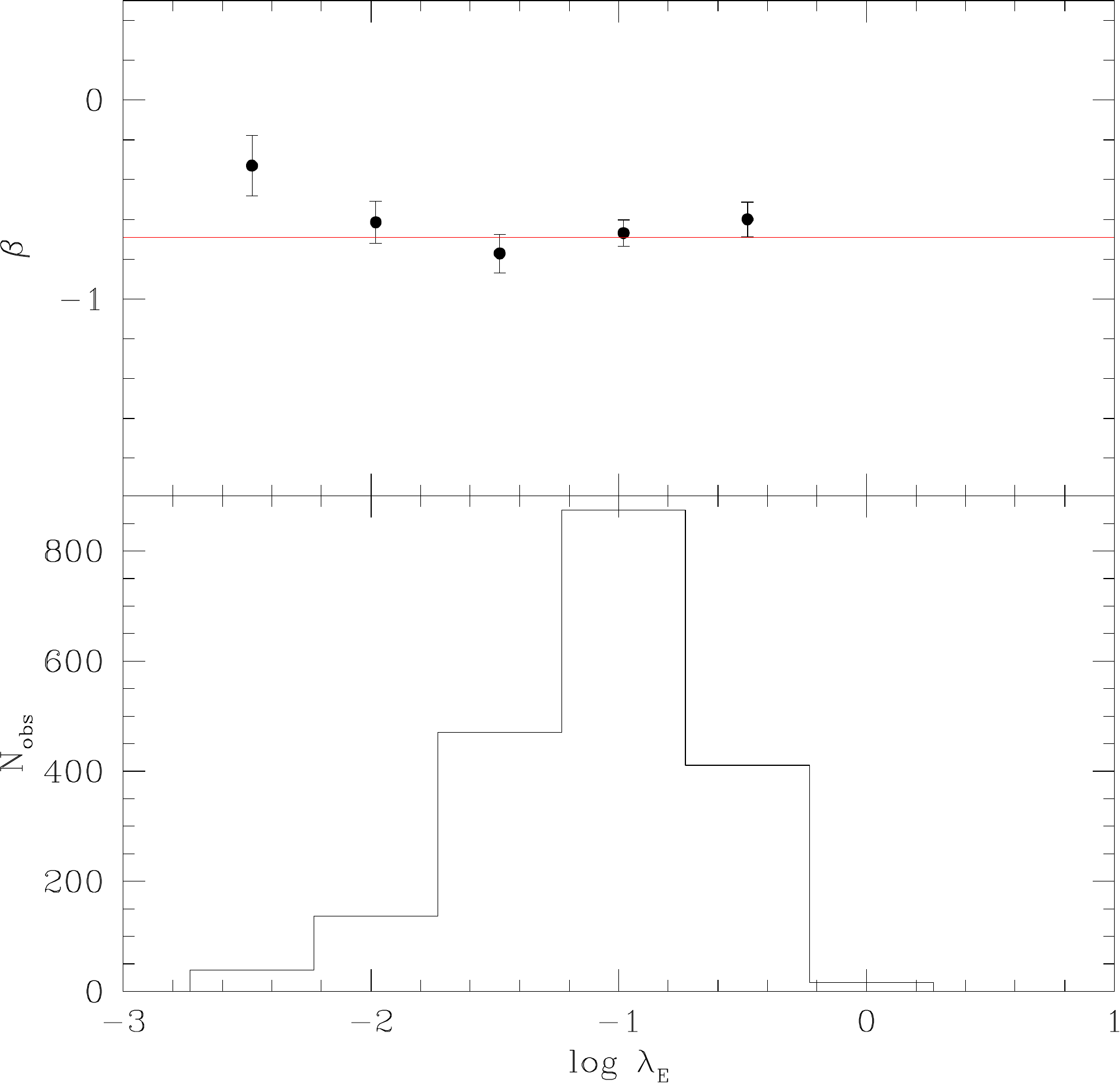}
\caption{\textit{Top}: spectral variability parameter $\beta$ versus Eddington ratio $\lambda_E$ (top). There is no clear trend for $\lambda_E\gtrsim-2$. Even though the number of points for the first bin is scarce, the absolute value of $\beta$ for very low Eddington ratio $\lambda_E\lesssim-2$ is smaller, suggesting that for even lower Eddington ratios it might become positive, transitioning to harder when brighter trend as found by \citet{connolly16}. \textit{Bottom}: histogram of Eddington ratio is also shown.}
\label{fig:eddratio}
\end{figure}

\noindent We also divided the sample into bins of Eddington ratio $\lambda_E=L_{bol}/L_{Edd}$ of $\delta\log\lambda_E=0.5$ amplitude. As shown in Fig.~\ref{fig:eddratio}, $\beta$ does not depend on $\lambda_E$ either for $-2\lesssim\lambda_E\lesssim0$, while it shows a weak trend toward smaller absolute values for $\lambda_E\lesssim-2$, suggesting a possible agreement with \citet{connolly16}. In fact, according to their work, there might be a transition to harder when brighter trend ($\beta>0$) for very low Eddington ratio sources. The trend in our data is not strong enough, given the small number of observations contributing to small $\lambda_E$.

\begin{table*}
\centering
\begin {tabular}{c c c c c c c c c}
\hline
\texttt{IAUNAME} & Serial number & $\beta\pm\sigma(\beta)$ & $\langle F_S \rangle$ & $N_{obs}$ & $r$ & $p$ & $z$ & $\log L_X$\\
\hline
3XMM J095847.8+690532 & $97$ & $-1.57\pm0.28$ & $7.32\cdot10^{-14}$ & $13$ & $0.861$ & $7.9\cdot10^{-5}$ & $1.288$ & $45.02$\\
3XMM J100043.1+020637 & $117$ & $-1.17\pm0.15$ & $3.80\cdot10^{-14}$ & $11$ & $0.932$ & $1.5\cdot10^{-5}$ & $0.360$ & $43.46$\\
3XMM J121753.0+294304 & $228$ & $-3.54\pm0.54$ & $3.34\cdot10^{-14}$ & $8$ & $0.938$ & $2.9\cdot10^{-4}$ & $1.647$ & $44.88$\\
3XMM J122931.2+015247 & $256$ & $-1.43\pm0.39$ & $1.70\cdot10^{-14}$ & $24$ & $0.613$ & $7.1\cdot10^{-4}$ & $0.770$ & $43.87$\\
3XMM J123759.5+621102 & $269$ & $-0.81\pm0.10$ & $5.63\cdot10^{-14}$ & $7$ & $0.964$ & $2.4\cdot10^{-4}$ & $0.910$ & $44.57$\\
3XMM J013943.1+061254 & $532$ & $-0.80\pm0.06$ & $5.82\cdot10^{-14}$ & $5$ & $0.992$ & $3.9\cdot10^{-4}$ & $0.678$ & $44.33$\\
3XMM J115535.8+232723 & $2066$ & $-0.80\pm0.06$ & $9.24\cdot10^{-14}$ & $5$ & $0.991$ & $4.6\cdot10^{-4}$ & $0.136$ & $42.95$\\
3XMM J121834.6+293453 & $2125$ & $-1.38\pm0.11$ & $1.92\cdot10^{-14}$ & $8$ & $0.980$ & $9.3\cdot10^{-6}$ & $1.843$ & $44.81$\\
\hline
\end{tabular}
\caption{Single-source results are shown in this table. To identify each source, we have reported both the 3XMM-DR5 catalogue entry \texttt{IAUNAME} in the first column and the serial number in the MEXSAS catalogue that labels each source, which is shown in the second column. For each source, the spectral variability parameter $\beta$ and its error, the number of epochs $N_{obs}$, the correlation coefficient $r$, the probability of finding such correlation by chance $p$, the redshift $z$ and the X-ray luminosity $\log L_X$ are shown. Also shown is the average value of the flux $\langle F_S \rangle,$ from which we can see that all of the most correlated sources come from the two brightest flux bins (see Tab.~\ref{tab:betaflux}).}
\label{tab:singlesource}
\end{table*}

\subsubsection{BALQSOs and radio-loud sources}
\label{sec:balradio}
As briefly reported in Sec.~\ref{sec:dataset}, Broad Absorpion Line quasars (BALQSO) and radio-loud sources are also present in our sample. Both classes might differ in spectral variability properties from the overall trend, due, respectively, to X-ray absorption typically associated with the UV BAL outflows \citep[e.g. ][]{brandt00} and to the enhanced X-ray emission associated with their jets \citep[e.g. ][]{worrall87}.

To quantify the number of BAL sources present in our catalogue, we cross-matched it with the already mentioned catalogue by \citet{shen11}, which lists a flag to recognise BALQSO based on the catalogue by \citet{gibson09}. We find 31 observations of 14 sources in the DR7Q subset of our catalogue. As for the DR12Q subset of our catalogue, the quasar catalogue itself lists the BALQSO flag \citep{paris16} and we obtain 171 observations of 63 sources. Removing 13 observations of 6 sources that are found in both DR7Q and DR12Q, we have a total of 189 observations of 71 unique sources for the merged BALQSO list.

The spectral variability parameter of this subset was then computed and we find $\beta=-0.64\pm0.16$, with correlation coefficient $r=0.285$ and a probability of finding such correlation by chance given by $p(>r)\simeq9\cdot10^{-5}$. This estimate is not dissimilar from the ensemble average value of the spectral variability parameter and, therefore, the BALQSOs do not affect the general ensemble analysis.\\

We then cross-matched the \citet{shen11} and the MEXSAS again in order to find radio-loud sources. We identified 169 observations of 56 unique sources with available radio data, with 137 observations of 47 unique sources that have radio loudness parameter $R>10$, defined as

\begin{equation}
R=\frac{F_{6~\textrm{cm}}}{F_{2500~\AA}},
\end{equation}

\noindent after \citet{jiang07}, that is, they are radio-loud sources. This catalogue only lists data for the sources from the SDSS-DR7Q.

Therefore, we identified the FIRST flux \citep{becker95} at $\lambda=20$~cm in the SDSS-DR12Q catalogue and, in order to have comparable results with \citet{shen11}, we computed the radio flux at rest-frame wavelength $\lambda=6$~cm using $F_\nu\propto\nu^{\alpha}$. Following \citet{gibson08} and \citet{vagnetti10}, we assume a standard $\alpha_{radio}=-0.8$ value for all the sources. We also identified the $ugriz$ fluxes in the SDSS  \citep{fukugita96} and computed the flux at $2500\AA$ by interpolating between the nearest SDSS filters in the rest-frame, or extrapolating in the cases when all the filters lie on the same side with respect to $2500\AA$, adopting in such cases $\alpha_{UV}=-0.46$, following \citet{vandenberk01}.

Then, we computed the radio-loudness parameter also following \citet{jiang07}. We found radio data for 244 observations of 80 sources in the SDSS-DR12Q catalogue, which makes a total of 126 sources at 384 epochs if we include the \citet{shen11} data for the DR7 sources and exclude nine sources that are available in both catalogues, for which we chose to adopt the radio-loudness as computed by us. In this sample, $R$ ranges from 0.3 up to $5\cdot10^{4}$. Most of these sources (114) are radio-loud ($R>10$) and we find a spectral variability parameter $\beta=-0.38\pm0.10$ for this sample. Compared to the result obtained for all the sources (see Eq.~\ref{eq:betaensemble}), this means that, on average, the spectra of radio-loud sources are less variable than the rest. A similar behaviour is also seen in the optical/UV bands \citep{vagnetti03}.

\subsection{Individual sources}
\label{sec:indisources}
As shown in Sect.~\ref{sec:subsamples}, the ensemble spectral variability parameter $\beta$ does not appear to change with the distinctive features of the quasar, such as black-hole mass, redshift, Eddington ratio or X-ray luminosity. However, as is clear from Figs.~\ref{fig:betaredshift}, \ref{fig:betalx}, \ref{fig:bhmass} and \ref{fig:eddratio}, some kind of deviation from the ensemble value is obviously allowed. For some bins of X-ray luminosity, we even find $\beta>0$, hinting at a harder when brighter behaviour for the sources present in those bins, opposite to the general trend of the whole sample (see Fig.~\ref{fig:betalx}).

This means that single sources as well could deviate even significantly from the general trend of the ensemble. To study this, we computed $\beta$ for each source with $N_{obs}>3$. We specify that most of these $\beta$ values are not significant, because they are either not correlated or the fits are performed on too few points to be acceptable.

\begin{figure}
\centering
\includegraphics[width=\columnwidth]{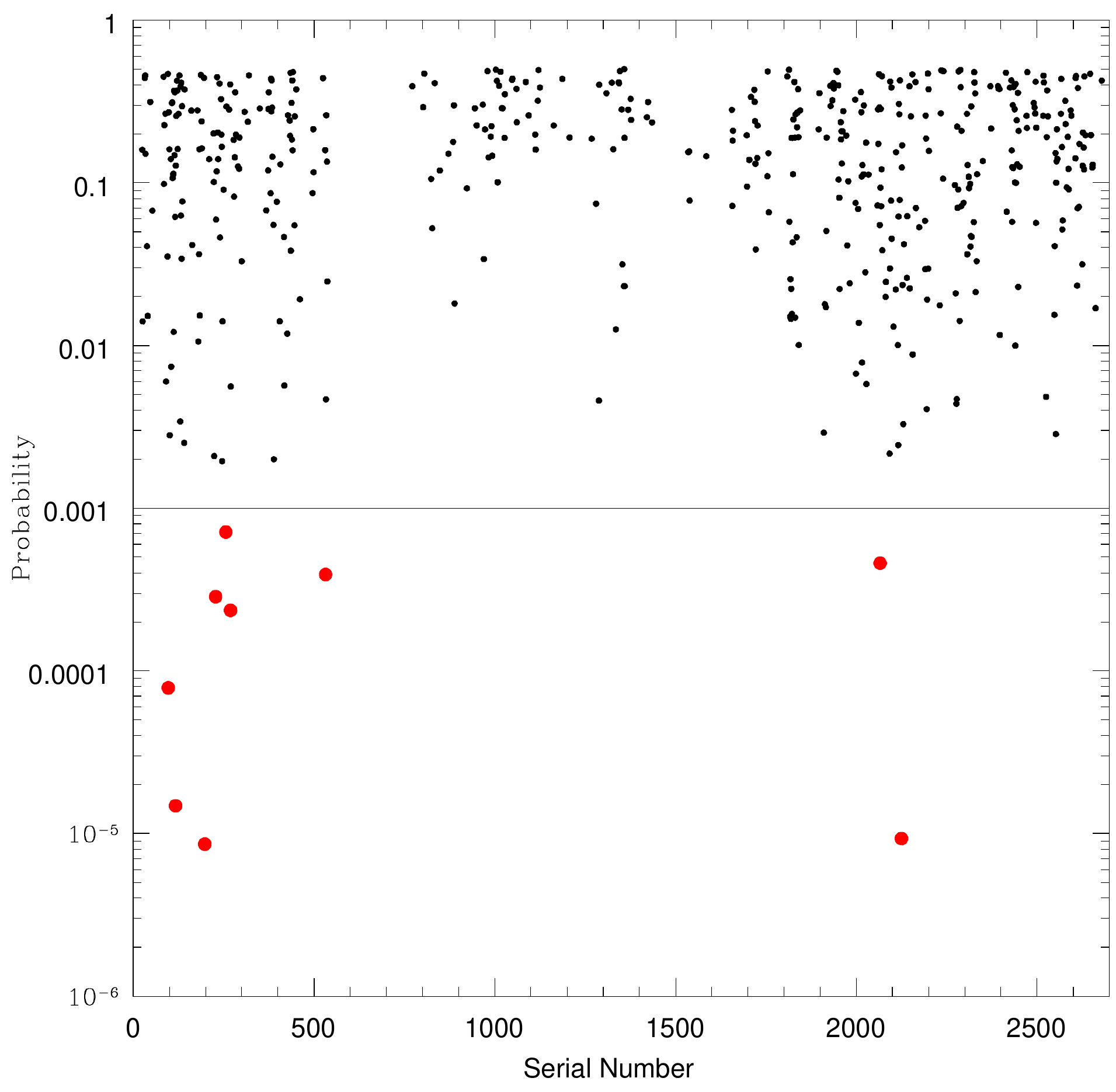}
\caption{Probability of finding $\Gamma-\log F_S$ correlation by chance. The most significantly correlated sources are chosen by having probability lower than $10^{-3}$ (horizontal line). The nine selected sources are shown as red circles. One of these sources (198, also known as PG 1114+445) was discarded due to the well-known presence of a warm absorber, reducing the number of analysed sources to eight.}
\label{fig:probsingle}
\end{figure}

Many of the sources listed have a very low individual correlation between their own fluxes $F_S$ and spectral indices $\Gamma$, which means that they are mildly variable in spectrum or have too few observations to tell. The probability of finding a significant $\Gamma-\log F$ correlation by chance is plotted in Fig.~\ref{fig:probsingle} for each source. To estimate this correlation probability, given the small number of points involved, we perform a Student's t-test \citep{bevington69}, which computes the integral probability of the null hypothesis $p(>r,N)$ for a given correlation coefficient $r$ with $N$ pairs of data points. We selected those sources for which the probability of finding a casual correlation is very small ($p\leq10^{-3}$), in order to find the most confident results. We have therefore selected nine sources, eight of which are listed in Table \ref{tab:singlesource}. Indeed, after visual inspection of the available X-ray spectra, we excluded source number 198, that clearly shows the presence of a warm absorber. This source can be identified with PG 1114+445, a well-known source whose warm absorber has already been studied in the past \citep[e.g. ][]{ashton04}. In the ensemble study, we estimated that the presence of additional components, such as a warm absorber, only affect $10\%$ of the sources, and therefore can be neglected. However, the presence of these features in a single source study requires a more detailed analysis, which is beyond the scope of this paper. Therefore, we only analysed the eight individual AGNs listed in Table \ref{tab:singlesource}. We note that these sources are distributed in a wide range of X-ray luminosities and redshift.

\begin{figure*}
\centering
\includegraphics[width=0.75\textwidth]{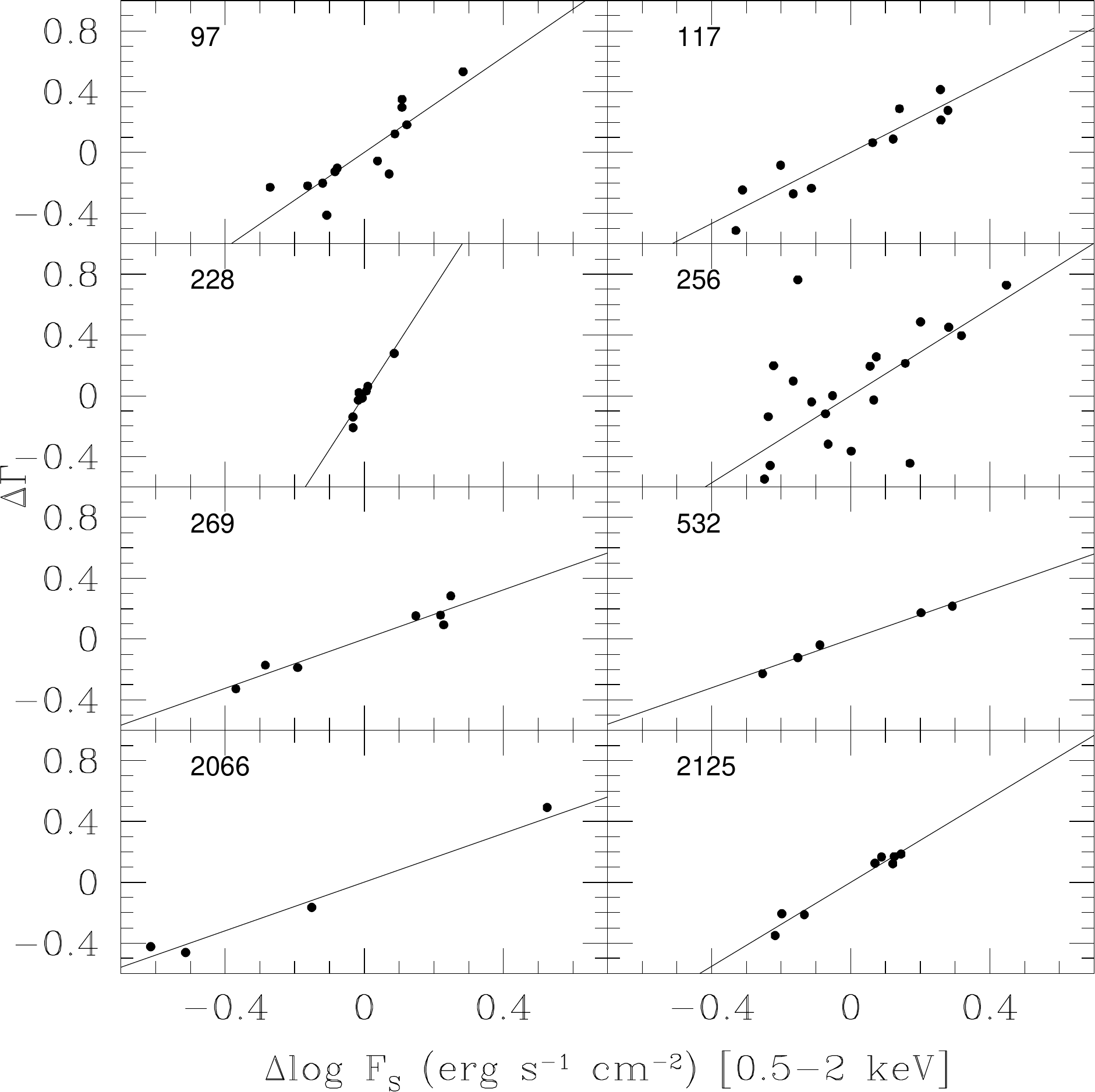}
\caption{We report here the $\Delta\Gamma - \Delta\log F_S$ plots for eight sources with high correlation and low ($p\leq10^{-3}$) probability of finding such correlation by chance. The slopes are different from source to source, ranging from $\beta=-3.54$ to $\beta=-0.80$, suggesting that even though the mechanism behind spectral variability is likely the same, there is a certain range among the sources. The MEXSAS serial numbers are reported in each panel, see also Table~\ref{tab:singlesource}.}
\label{fig:indisources}
\end{figure*}

All of them show a softer when brighter behaviour, although to a varying extent, since $\beta$ ranges from $\beta=-0.80\pm0.06$ for sources numbered 532 and 2066, to the extreme value of $\beta=-3.54\pm0.54$ for the source numbered 228 in the MEXSAS catalogue (see Fig.~\ref{fig:indisources}). All of the sources belong to the two highest-flux bins, confirming that the brightest sources are the most correlated ones.

\section{Summary and discussion}
\label{sec:discussion}

We have found an ensemble softer when brighter behaviour for a sample of 2,700 quasars by means of the spectral variability parameter $\beta$, first introduced by \citet{trevese02}. A negative value ($\beta=-0.69\pm0.03$) was found, which means that a positive slope between $\Delta\Gamma$ and $\Delta\log F$ is present. This positive slope means that, on average, a softer when brighter trend is present for our sample. As can be seen in Figs.~\ref{fig:dgdlogf1} and \ref{fig:dgdlogf2}, the errors in $\Delta\Gamma$, which are higher for the sources with lower average flux, give rise to a wider dispersion of the points in the $\Delta\Gamma - \Delta\log F_S$ plane, causing a steeper slope of the linear fit, meaning a larger absolute value of $\beta$. Even considering only the sources with the best spectrum (the ones in common with the XMMFITCAT catalogue released by \cite{corral15}), we still obtain a negative $\beta$, slightly smaller in absolute value, $\beta=-0.43\pm0.04$. Dividing our sample into average flux bins, the brightest sources give $\beta=-0.54\pm0.03$, still confirming a softer when brighter trend.

This result should be viewed in the context of the existing models for the X-ray emission and spectral variability engines.

The current paradigm for X-ray emission involves the presence of two mutually interactive components: a cold, optically thick  phase, that is, the accretion disk, that provides input soft UV photons, and an optically thin phase, that is, a hot electron plasma, commonly known as the corona \citep{liang79, haardt91}. The soft photons that are emitted by the accretion disk are up-Comptonised by the corona in the X-ray band with a power-law spectrum. \citet{nandra01} found that the X-ray changes of the spectral index respond to the UV emission variations, when studying the spectral variability of the nearby source NGC 7469. Part of the X-ray radiation emitted by the corona gets reflected by the disk and torus \citep{magdziarz95, mchardy98, shih02}. 

The observed softer when brighter trend might be due to the fact that the soft part of the spectrum is variable in amplitude but not in photon index, while the reflection component is approximately constant. In that case, the variations of the observed photon index $\Gamma_{obs}$ are due to the combination of these two spectral components, but not to changes of the intrinsic photon index $\Gamma_{int}$ \citep[see e.g. ][]{taylor03, ponti06, miniutti07}.

Another option is that the soft component itself is already variable in spectrum, and that the variations of the observed spectral index $\Gamma_{obs}$ correspond to changes of the intrinsic spectral index $\Gamma_{int}$, which increases with decreasing coronal temperature \citep{haardt97}.

We also tried to identify different trends for different source features, finding no clear trend between $\beta$ and the redshift $z$, the X-ray luminosity $L_X$, the black-hole mass $M_{BH}$ or the Eddington ratio $\lambda_E$. This suggests that the main mechanism that generates the observed spectral variability trend is likely the same for all sources and probably not due to the physical characteristics of the source. Even though the spectral variability parameter and these quantities are not correlated, there is some deviation from the average value of $\beta$ on different bins of $z$, $L_X$, $M_{BH}$ and $\lambda_E$. We have also computed $\beta$ for a subsample of $114$ radio-loud sources, finding a smaller spectral variability with respect to the general trend. A subset of 71 BALQSOs was also considered, but the spectral variability parameter does not sensibly differ from the ensemble value.

Individual $\beta$s were also computed for a restricted set of eight sources, which display the most significant $\Gamma-\log F$ correlations. For these sources, $\beta$ ranges from $\beta=-3.54\pm0.54$ to $\beta=-0.80\pm0.06$. It is not clear why the spectral variability parameter, although still maintaining the softer when brighter trend in all cases, is quantitatively different from source to source. One possibility is that they have different orientations and that the spectral variability could be dependent on the angle of view $i$. Another possibility is a dependence on the black-hole spin, about which there is no available information for these sources. The presence of hidden spectral features, such as soft excess and/or warm absorbers, could also alter the observed value of $\beta$. Alternatively, the process could be entirely stochastic, and therefore not correlated to any of the quantities mentioned above.

Future observations of these objects might hopefully shed some light on the topic. For instance, in the near future the mission eROSITA \citep{merloni12} is scheduled to launch. This mission is primarily dedicated to investigating the mass function of galaxy clusters, and  during the first four years will perform all-sky searches in order to find $10^4-10^5$ such objects. It is estimated \citep{predehl10} that it will be able to detect $\sim10^6-10^7$ AGNs with redshifts up to $z\sim7-8$, with an average of eight flux measurements for each source over the four years \citep{merloni12}. One such vast catalogue would significantly improve the statistical significance of such variability studies.

\begin{acknowledgements}
We acknowledge funding from PRIN/MIUR-2010 award 2010NHBSBE. We thank Stefano Bianchi, Sunil Chandra, Paola Marziani and Maurizio Paolillo for useful discussions. We would like to thank the referee for giving constructive comments that helped improving the quality of the paper. This research has made use of data obtained from the 3XMM XMM-Newton serendipitous source catalogue compiled by the 10 institutes of the XMM-Newton Survey Science Centre selected by ESA. Funding for SDSS-III has been provided by the Alfred P. Sloan Foundation, the Participating Institutions, the National Science Foundation, and the U.S. Department of Energy Office of Science. The SDSS-III web site is http://www.sdss3.org/. SDSS-III is managed by the Astrophysical Research Consortium for the Participating Institutions of the SDSS-III Collaboration including the University of Arizona, the Brazilian Participation Group, Brookhaven National Laboratory, Carnegie Mellon University, University of Florida, the French Participation Group, the German Participation Group, Harvard University, the Instituto de Astrofisica de Canarias, the Michigan State/Notre Dame/JINA Participation Group, Johns Hopkins University, Lawrence Berkeley National Laboratory, Max Planck Institute for Astrophysics, Max Planck Institute for Extraterrestrial Physics, New Mexico State University, New York University, Ohio State University, Pennsylvania State University, University of Portsmouth, Princeton University, the Spanish Participation Group, University of Tokyo, University of Utah, Vanderbilt University, University of Virginia, University of Washington, and Yale University.
\end{acknowledgements}

\end{document}